\documentclass[acmlarge,nonacm]{acmart}

\AtBeginDocument{%
  \providecommand\BibTeX{{%
    \normalfont B\kern-0.5em{\scshape i\kern-0.25em b}\kern-0.8em\TeX}}}

\usepackage{caption}
\usepackage{subcaption}

\setcopyright{acmcopyright}
\copyrightyear{2018}
\acmYear{2018}
\acmDOI{10.1145/1122445.1122456}

\acmJournal{TOG}
\acmVolume{37}
\acmNumber{4}
\acmArticle{111}
\acmMonth{8}

\begin{document}

\title{Eliciting Gestures for Novel Note-taking Interactions}

\author{Katy Ilonka Gero}
\authornote{Research performed while an intern at Google Research.}
\email{katy@cs.columbia.edu}
\affiliation{%
  \institution{Columbia University}
  \city{New York City}
  \state{New York}
  \country{USA}
}

\author{Lydia B. Chilton}
\email{chilton@cs.columbia.edu}
\affiliation{%
  \institution{Columbia University}
  \city{New York City}
  \state{New York}
  \country{USA}
}   

\author{Chris Melancon}
\email{cmelancon@google.com}
\affiliation{%
  \institution{Google Research}
  \state{Oregon}
  \country{USA}
}

\author{Mike Cleron}
\email{mcleron@google.com}
\affiliation{%
  \institution{Google Research}
  \city{California}
  \country{USA}}

\renewcommand{\shortauthors}{Gero, et al.}

\begin{abstract}
  Handwriting recognition is improving in leaps and bounds, and this opens up new opportunities for stylus-based interactions. In particular, note-taking applications can become a more intelligent user interface, incorporating new features like autocomplete and integrated search. In this work we ran a gesture elicitation study, asking 21 participants to imagine how they would interact with an imaginary, intelligent note-taking application. We report agreement on the elicited gestures, finding that while existing common interactions are prevalent (like double taps and long presses) a number of more novel interactions (like dragging selected items to hotspots or using annotations) were also well-represented. We discuss the mental models participants drew on when explaining their gestures and what kind of feedback users might need to move to more stylus-centric interactions.
\end{abstract}




\maketitle

\section{Introduction}

As handwriting and hand-drawn figure recognition improves \cite{dutta2018improving, ptucha2019intelligent, yousef2020accurate}, we see new opportunities for stylus-based interactions. Features like autocomplete and integrated search, previously confined to text-based computing environments, may soon be implemented in handwriting-based environments. People report that using a pen still affords the most high precision, ``in the flow'' interactions, even in the face of the variety of computing mediums available (including digital pens) \cite{riche2017we}, and so in this work we consider how note-taking applications can become a more intelligent user interface, automatically responding to users' pen actions without requiring as many modes or menus. 
We envision future note-taking applications to be more like the word processor in terms of how it departed from the typewriter: future note-taking applications will make interacting with handwritten, nonlinear notes -- not just words, but also diagrams and sketches -- as easy as interacting with typed text is currently in modern word processors.

While studying gestures for surface computing has a long history in human computer interaction \cite{liao2005papiercraft,wobbrock2009user, zhai2012foundational}, less attention has been paid to gestural interactions for general, stylus-based note-taking activity. Current work tends to focus on specific applications that are complementary to keyboard-centric digital activities, like handwritten annotations on typed documents \cite{talkad2018replicating, yoon2013texttearing, yoon2014richreview, romat2019spaceink}, or drawing and note-taking as part of data analysis \cite{kim2019inking, xia2018dataink}. But general purpose, handwritten notes remain a mainstay in our lives, despite all the computing mediums available \cite{riche2017we}.

Our work focuses on entirely handwritten note-taking as a highly impactful use case of surface computing. Handwritten notes, and the physical embodiment that they afford, have been shown to improve the retention of lecture material \cite{mueller2014pen, morehead2019much}, be a critical component of the design process \cite{suwa2002external}, as well as allow for the benefits of nonlinear thinking \cite{makany2009optimising}. 
Making handwritten notes more interactive, by giving them the editing capabilities that computers allow, may allow handwritten notes to take on the power currently only seen when working with 3D, physical objects, which has been shown to increase people's causal reasoning \cite{vallee2015interactivity} and improve their ability to solve insight problems \cite{vallee2016insight}.

In this work, we seek to understand how users envision interacting with an intelligent note-taking application where handwritten notes are a first-order object and users are able to interact with `digital ink' without always having to revert to menu options or a keyboard. We ran an elicitation study, asking 21 participants to imagine how they would perform eight unique actions. We found that while some actions have a very high level of agreement, others are quite distributed, with participants drawing on a variety of mental models to imagine how to perform the novel action.
  
Our contributions are:

\begin{itemize}
    \item elicited gestures for eight note-taking actions, including autocomplete actions, and their level of agreement;
    \item a description of the mental models that participants drew on, including the word processor model; and
    \item a discussion of how feedback might guide users' understanding of digital ink affordances.
\end{itemize}

\section{Related Works}

\subsection{Gesture Elicitation Studies}

Gestures for surface computing have a long history in HCI, with foundational issues like motor control complexity, visual and auditory feedback, and memorability being well studied
\cite{zhai2012foundational}. Elicitation studies for gestures, or `user-defined gestures', are often used to understand how people currently think about surfacing computing while simultaneously aiding in the design of gestures for novel interactions \cite{wobbrock2009user, talkad2018replicating}. \cite{nacenta2013memorability} found that user-defined gestures are easier to remember and that participants prefer them, though in commercial products gestures tend to be pre-defined (though informed by much user testing, and users may have options for customization).

One critique of the gesture elicitation study methodology is often termed `legacy bias', where participants' proposed gestures are heavily biased by their experience with existing computing interactions. Several methods have been proposed to reduce legacy bias, including having participants produce more than one gesture per referent \cite{morris2014reducing}. Though this technique has seen mixed results in reducing legacy bias \cite{williams2020cost}, we implement a version of it in this study that also results in qualitative data about participants' mental models.

\subsection{Digital Note-taking}

As the use of touchscreen devices increases, primarily smartphones but increasingly larger tablets, surface computing interactions have become much more common, opening up research for new applications like collaborative document reviewing \cite{yoon2014richreview, yoon2016richreview++} and creating data visualizations \cite{kim2019datatoon, kim2019inking}. Many of these areas use elicitation studies to understand how users want and expect to perform new actions.
Simultaneously, improvements in not just handwriting recognition \cite{dutta2018improving, ptucha2019intelligent, yousef2020accurate} but also handwriting generation \cite{aksan2018deepwriting, fogel2020scrabblegan} have opened up new opportunities for handwriting-based surface computing. Handwritten notes are an incredibly varied and important medium. \cite{riche2017we} did an extensive diary study of analog pen use, and reported 9 affordances of pens, including externalizing thoughts (often to remember things), producing high-fidelity marks (indicating the precision and fluidity of expression pens provide), and automatic usage (people often feel like writing with a pen keeps them ``in the flow''). All of these speak to the importance people still find in pen-based expression over all the computing mediums available to them. 

Though the adoption of styluses and their related applications is slow, we expect to see an increase in their usage as technology improves and their affordances approach those found in analog pens.
For instance, features previously only seen in text-based environments, like autocomplete and integrated search, can now be implemented in handwriting-based environments. In this work we study how users imagine interacting with such an intelligent user interface, in order to aid in the improvements of these technologies.

\section{Methodology}

\subsection{Study Design}

\begin{figure}
  \centering
  \includegraphics[width=.95\linewidth]{
  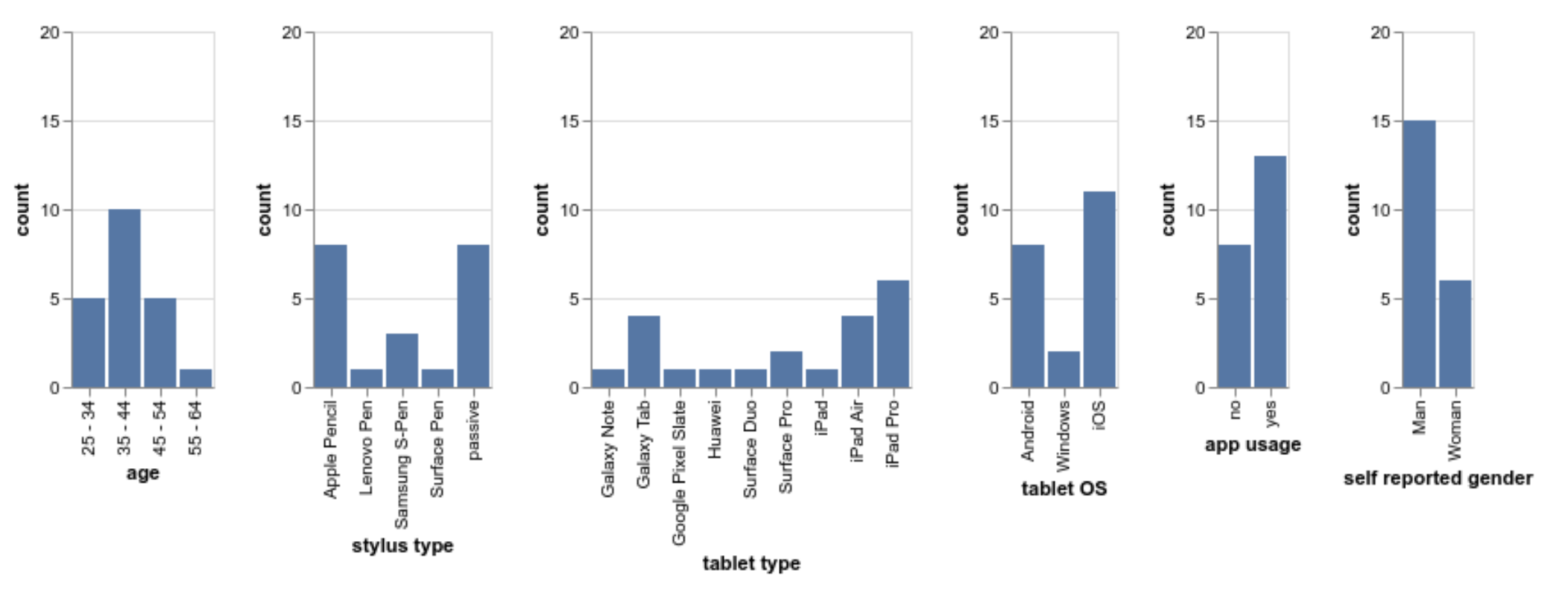}
  \caption{Histograms of all the demographic data we collected about the 21 participants in our study. We are able to recruit a wide range of tablet types and stylus types. Note that `app usage' refers to whether or not a participant reported using a stylus-based note taking app.}
  \Description{A woman and a girl in white dresses sit in an open car.}
  \label{fig:demo}
\end{figure}

We recruited 21 participants for the study from the USA, and the entire study was run in English. All participants had to have used a tablet for at least 6 months and own a stylus. We recruited participants from a variety of age ranges, as well as a variety of tablet types and operating systems. The demographics of our participants can be found in \autoref{fig:demo}.

\begin{figure}
  \centering
  \includegraphics[width=.75\linewidth]{
  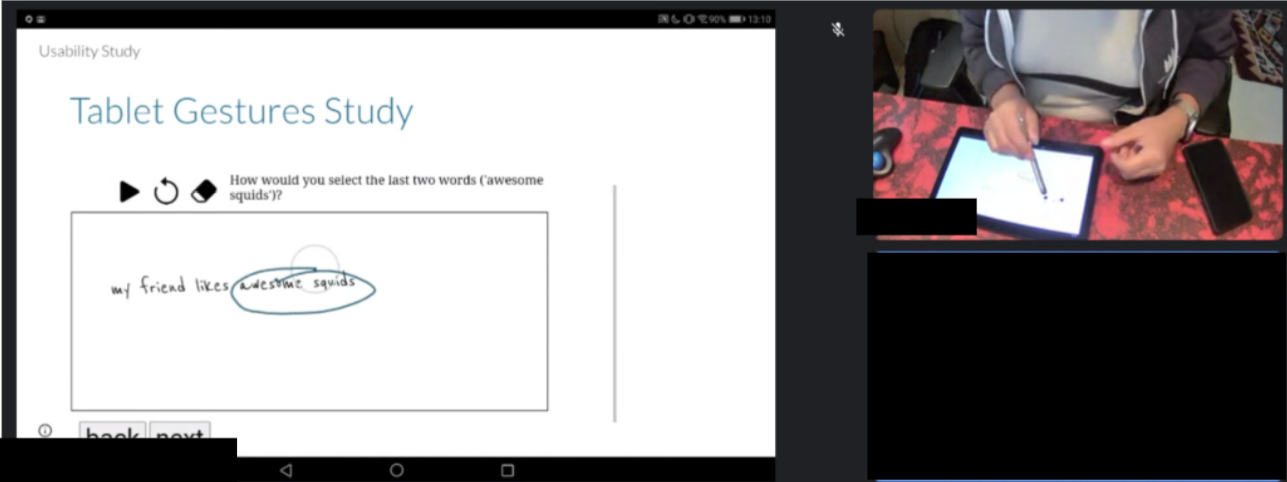}
  \caption{A screenshot of the study set up as it appeared to the facilitator. All personally identifiable information has been redacted. On the left the participant is screen sharing from their tablet, and on the right we had a camera pointing at their tablet such that we could see their hands.}
  \Description{A woman and a girl in white dresses sit in an open car.}
  \label{fig:setup}
\end{figure}

Due to the covid-19 pandemic, our study took place remotely. Participants joined a video conference from a computer and from their tablet. They shared the screen of their tablet, and pointed the webcam of their computer at their tablet such that we were able to see their hands when they interacted with the tablet. An example screenshot of the study setup, with all personally identifiable information removed, can be found in \autoref{fig:setup}.

Participants were told that the study was about how people wanted to interact with tablet-based note-taking applications, and was not a test of their knowledge of existing interactions. They were told that they would be shown a series of still images that they should imagine are part of a note-taking application -- any text in these images should be assumed to have been handwritten. For each image they would be asked how they would take some action on the screen, and they would be asked to show both a stylus and a finger interaction. They were then taken through an example elicitation exercise, where they were shown some words on a screen and asked how they would move these words downward.  

When participants interacted with the image, touch events were recorded by leaving a trail -- in \autoref{fig:setup} you can see that as the participant circles the words, a mark of where their stylus has been is shown. The beginning of a touch event is indicated by a grey circle. Participants were told that this was part of the data collection process, and did not indicate that they were drawing on screen. The additional webcam pointed at their tablet allowed us to determine if they were using a stylus or a finger.

\begin{table}
  \begin{tabular}{ll}
    \toprule
    action&prompt\\
    \midrule
    select two words & How would you select these two words?\\
    select from phrase & How would you select the last two words ('awesome squids')?\\
    delete & How would you delete these two words?\\
    hide & How would you 'hide' or 'collapse' these bullet points?\\
    autocomplete accept & How would you accept the suggested text completion?\\
    autocomplete reject & How would you reject the suggested text completion?\\
    copy & How you could copy 'research squids' so you could paste it elsewhere?\\
    search & How would you search google for the last two words, 'sea snails'?\\
  \bottomrule
\end{tabular}
\caption{Actions and the text prompt provided to participants. Each action had a still image associated with it -- `select two words' showed just two words on the screen, while `select from phrase' showed those two words embedded in a short phrase. `hide' showed a phrase with some bullet points beneath it, `copy' showed a list of short phrases, and `search' showed a phrase.}
\label{tab:prompts}
\end{table}

Participants were then led through the actions listed in \autoref{tab:prompts} by the facilitator. Four of these actions (`select two words', `select from phrase', `copy', and `delete') were chosen as basic actions currently supported in most stylus-based note-taking applications.\footnote{In particular we considered GoodNotes, OneNote, Squid, and Notability.} The remaining four were selected as potential new features that improvements in handwriting recognition/generation and tablet-based software may be able to support in the future. 

Participants were encouraged to think aloud, which most participants did successfully. When a participant didn't explain or verbally think through a gesture that the facilitator found confusing or particularly unique, the facilitator would ask the participant to explain their reasoning or what they were thinking of when they did that gesture. Also if a participant exclusively indicated they would use a menu option or a keyboard, the facilitator would prompt them to provide an interaction that did not include menu options or keyboards. This enabled us to avoid issues in elicitation studies where participants rely solely on existing mental models, by first letting participants respond in their most intuitive way, and then asking them to continue thinking of other interactions. This methodology has been suggested by \cite{morris2014reducing}. Sometimes participants were unable to think of alternate interactions, even when prompted and given time, in which case the facilitator would allow them to move on. For these reasons, our data is less clean than in some elicitation studies -- for instance not all participants did the same number gestures. However, we found that this more conversational approach encouraged participants to think outside of menu interactions, and 
allowed us to gather rich qualitative data about participants' mental models.

\subsection{Data Coding}

The raw result of our study was 21 videos of participants responding to the action prompts listed in \autoref{tab:prompts}.
These videos were then split into individual action responses such that each clip contains a single participant responding to a single action prompt. The clips were then arranged into `supercuts' such that a researcher could view all of the responses to a single action prompt. We then proceeded to code the participants' gestures.

\begin{table}
  \begin{tabular}{ccc}
    \toprule
    action&num codes&annotator agreement\\
    \midrule
    select two words&8&.87\\
    select from phrase&8&.85\\
    delete&9&.84\\
    copy&10&.84\\
    hide&8&.94\\
    search&7&.78\\
    autocomplete accept&11&.83\\
    autocomplete reject&11& .75\\
  \bottomrule
\end{tabular}
\caption{The number of codes in each codebook, where there was one codebook per action, and the inter-annotator agreement reported as Cohen's Kappa. All prompts had a very high level of agreement.}
\label{tab:coded-prompts}
\end{table}

Some of the gestures also included a select action. For instance, for the `copy' action, participants often first selected the text and then did a gesture for copy. For this reason, some actions are also coded with a select code, in addition to the main code for that action. For each gesture we also code the pointer used, `stylus' `finger', or `both'.

Two researchers coded each action. First they each watched the relevant supercut alone, and took note of common gestures. Then they came together and developed a codebook which would allow them to code the most common gestures. Then they each watched the supercut again and coded each gesture. Finally an inter-annotator agreement score was calculated and disagreements were resolved with a third party. The inter-annotator agreement scores (Cohen's kappa) and number of gestures in each codebook can be found in \autoref{tab:coded-prompts}. Additionally, if an action also included a select gesture, all researchers used same `select' codebook for consistency.

Because different researchers created the codebooks for different actions, the codebooks had different levels of specificity. For instance, `copy' had three different codes that included some kind of tap gesture (`long press', `double tap', `triple tap') while search had none. In order to align the codes, two researchers created an `umbrella' codebook, where each code from all the codebooks was assigned a high-level `umbrella' code. This allowed us to compare all the actions while having more fine-grained coding data (the original codebooks and codes) to refer back to. 

\section{Results}

\begin{figure}
  \centering
  \includegraphics[width=.8\linewidth]{
  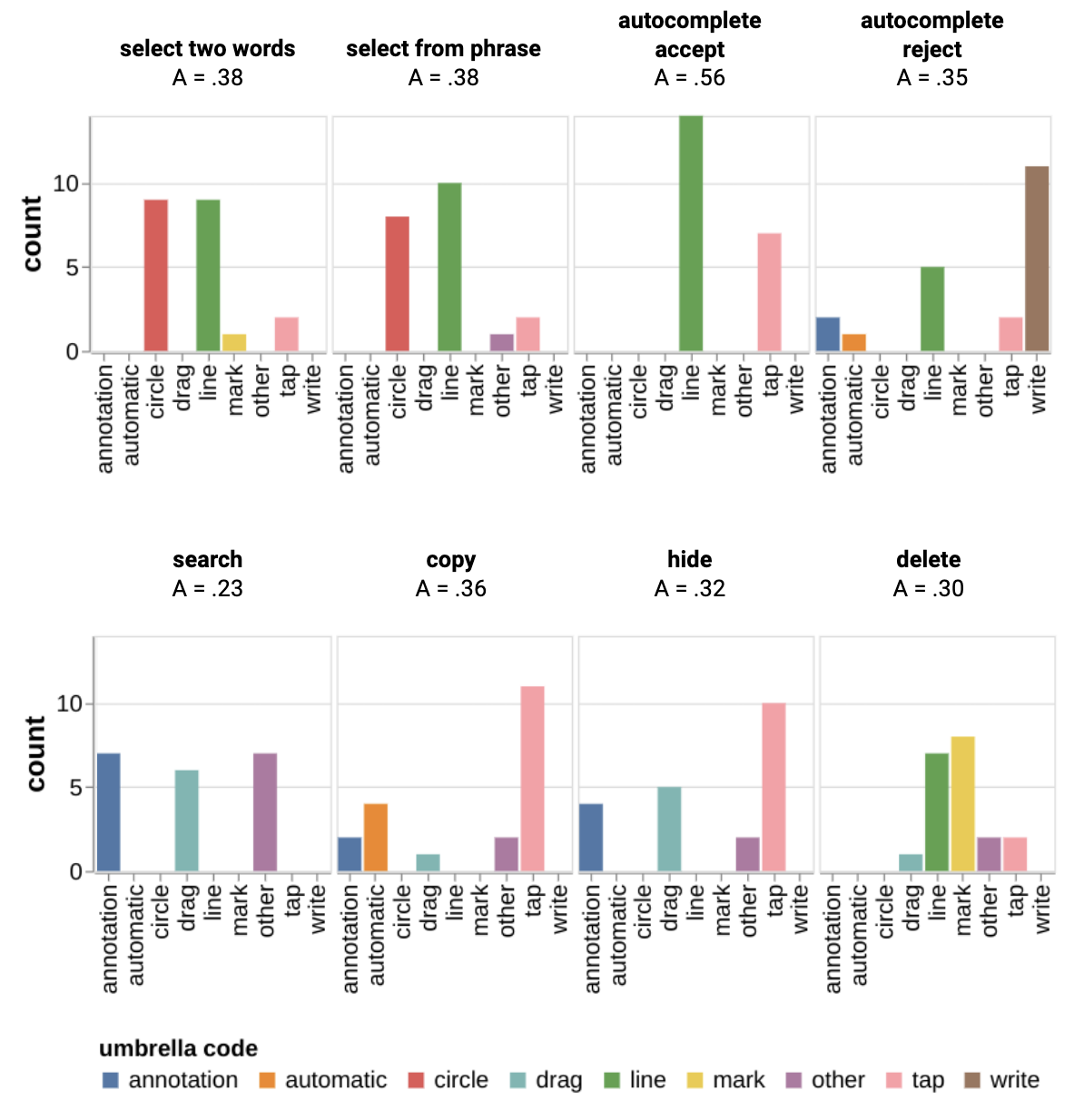}
  \caption{Here we show, for each action, the prevalence of the gesture types the participants performed, along with the level of agreement calculated for that action. While some actions have a clear consensus on a single gesture, for instance `line' for `autocomplete accept', many actions had two primary gestures, or for instance in the case of `search', an even distribution across three gestures.}
  \Description{A woman and a girl in white dresses sit in an open car.}
  \label{fig:prevalence}
\end{figure}
\begin{figure}
     \centering
     \begin{subfigure}{0.49\textwidth}
         \centering
         \includegraphics[width=\textwidth]{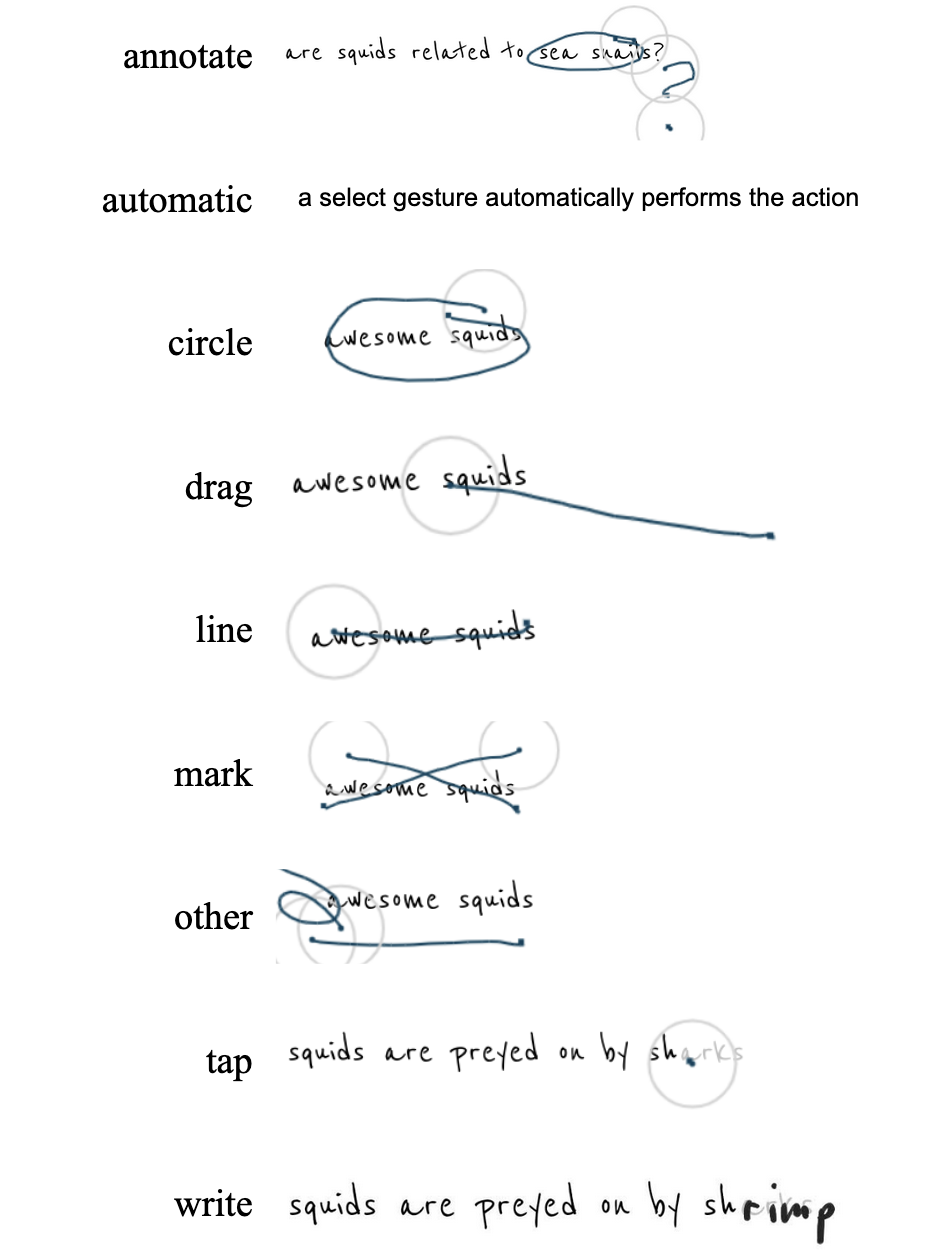}
         \caption{Illustration of each code we used to code all of the actions studied.}
         \label{fig:codes}
     \end{subfigure}
     \hfill
     \begin{subfigure}{0.49\textwidth}
         \centering
         \includegraphics[width=\textwidth]{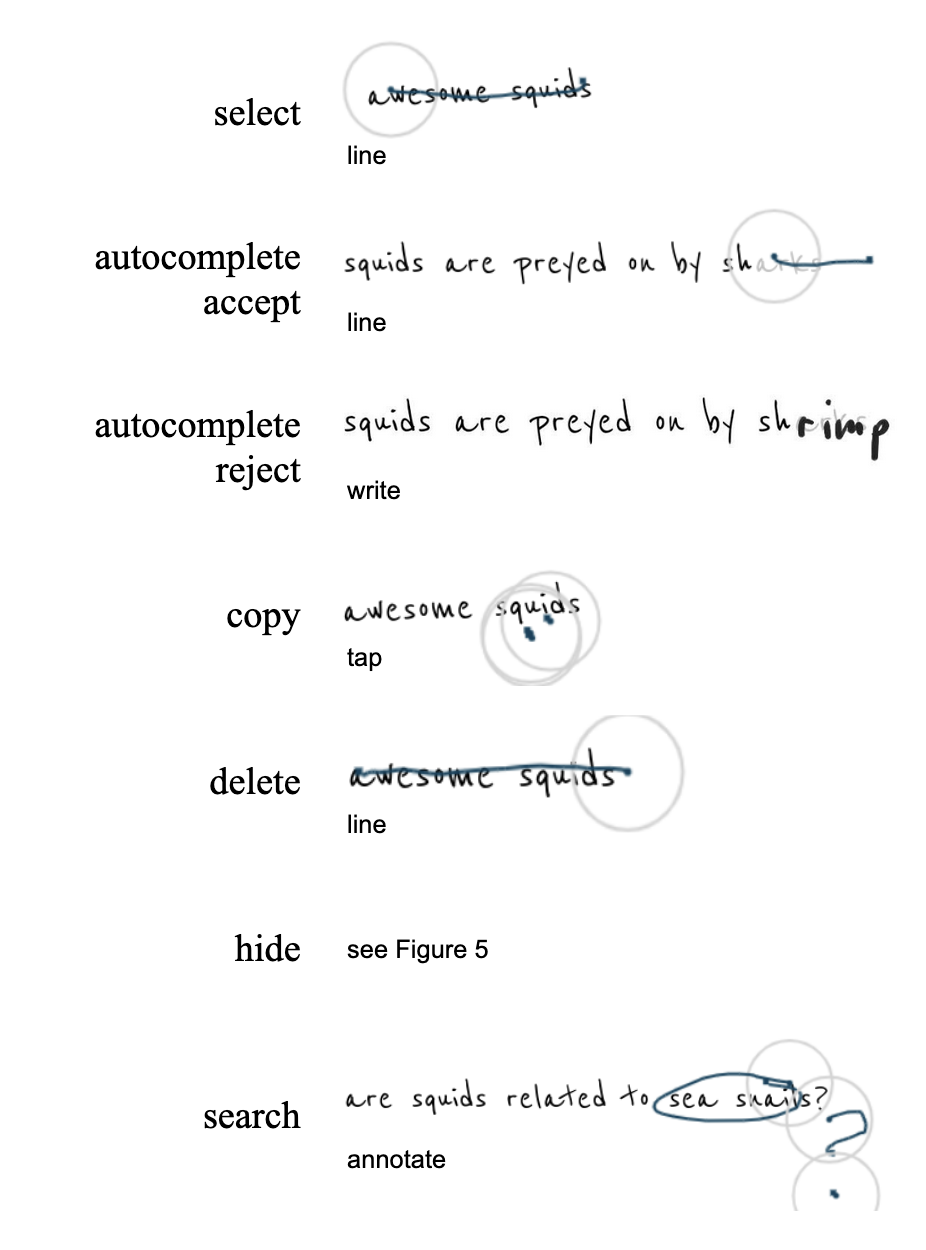}
         \caption{Most prevalent gesture performed by participants for each action studied.}
         \label{fig:action}
     \end{subfigure}
        \caption{Illustrations of various gestures. These illustrations are recreations of what we saw participants perform. Traces of a touch event are in the blue ink, and the gray circles represent the beginning of a touch event.}
        \label{fig:illustrations}
\end{figure}

\autoref{fig:prevalence} shows the prevalence of all gestures coded for each action. In particular, this figure shows the prevalence of the first non-menu gesture performed by participants. 
We report eight umbrella codes -- `annotation, `automatic', `circle', `drag', `line', `mark', `tap', `write', plus an `other' category. Examples for each code can be found in \autoref{fig:codes}, and an example of the most prevalent gesture for each action can be found in \autoref{fig:action}.

In \autoref{fig:prevalence} we also report the level of agreement, $A$, for each action. We calculate the agreement level as

\begin{equation*}
    A_{a} = \sum_{i} (\frac{|P_{i}|}{|P_{a}|})^2
\end{equation*}

where $a$ is a given action we studied, $P_{a}$ is the set of all gestures performed for this action, and $P_{i}$ is a set of gestures with identical codes. In the case of gestures in the `other' category, we count each gesture as in its own category.\footnote{e.g. If there are 20 gestures, and 12 are `line', 5 are `circle', and 3 are `other', then the agreement would be calculated as $(12/20)^2 + (5/20)^2 + (1/20)^2 + (1/20)^2 + (1/20)^2$.} This is in line with \cite{wobbrock2009user}. We calculate the agreement based on the umbrella codes. \cite{vatavu2019dissimilarity} notes that this agreement score is a function of subjective decisions by the researchers about how to group gestures, and we found that to be the case in our study as well. For instance, as reported the agreement score for `copy' is $0.36$ -- relatively high. But if we break the `tap' gestures into `long press', `double tap', `triple tap', the score drops to $0.165$. For this reason, we report the agreement in the context of the full prevalence data -- i.e. the histograms in \autoref{fig:prevalence} -- as well as more in-depth detail in the text of the paper.
We also report on how the select gesture correlates with different participant attributes, as well as qualitative data on the mental models that participants drew on when planning and explaining their gestures.

\subsection{Elicited Gestures}

\subsubsection{Select} 

The select gestures saw a relatively high level of an agreement, with most participants either selecting by circling the text or by swiping over the text in a straight line. Sometimes participants swiped through the text, for instance as one might expect if highlighting or striking through, while others swiped at a location that may be considered more like an underline. Some participants noted that when text was small it could be difficult for them to accurately strike through it, so underlines may have intended to be more like highlights. Since the difference between these was often ambiguous, and for this reason both as coded as `line'. 
While there were other gestures -- like tapping on words, or drawing the extents of an imagined selection box -- overall `select' saw a very high level of agreement, albeit across two different gestures. In \autoref{sec:select-correlates} we report on the select gestures used as part of other actions, and how consistent participants were in these gestures.

\subsubsection{Autocomplete Accept and Reject}

Autocomplete accept had the highest level of agreement of all the actions we studied, with the majority of participants drawing a line from left to right through the suggested text to indicate accept and minority of participants tapping on the suggested text to indicate accept. Participants talked about familiarity with this action, for instance when typing text messages or emails, and referred to the ways they would perform this in other applications.

Autocomplete reject had a reasonable consensus around `write', in which they would write over the suggested text. Again, participants referred to their experience with autocomplete in other contexts, where if they simply continued typing then the suggestion would disappear. However, we did see some participants use a line gesture, where the line would be drawn from right to left, which was also very common in the delete gestures.

\subsubsection{Search}

Search, in which participants were asked how they would perform a web search for text they had written from within the application, saw an even distribution across three different gesture categories, one of which was `other'. For this reason we consider search to have a fairly low level of consensus. The two common gestures that we did see were `annotation', where participants would draw something like a question mark or a `g' (for Google) to initiate a search on selected text, and `drag', where the participant would drag selected text to a special region on the page.
In the `other' category we saw gestures such as shaking the selected text or highlighting text with a special web search marker. But overall participants talked about how, while they were familiar with searching for text from various applications, they were unsure how to do this without moving to a browser.

\subsubsection{Copy}

The most prevalent gesture for performing a copy was some kind of tap gesture. Participants did a variety of different kinds of tap gestures, whether it be a double or triple tap, or a long press. Participants who used a long press noted that they may expect different actions to be performed at different lengths of time, for instance a short press indicating select while a one or two second press indicating copy, and an even longer press could perform some other actions. Several participants mentioned how they were aware that the various tap gestures were often overloaded, in that tap gestures already had a lot of actions associated with them. 
Another prevalent copy gesture was `automatic', in which any select gesture automatically resulted in a copy. This, combined with the prevalence of tap gestures, suggests that participants considered copy to be a common action that requires a simple, common gesture.

\subsubsection{Hide}

\begin{figure}
  \centering
  \includegraphics[width=.6\linewidth]{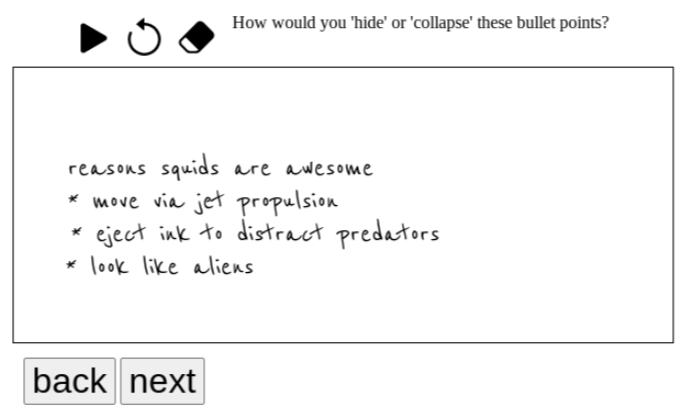}
  \caption{Prompt for the `hide' action as it was shown to participants. Participants often tapped on the top line (`header') or bullet points (`content') in order to perform the hide action.}
  \Description{A woman and a girl in white dresses sit in an open car.}
  \label{fig:hide}
\end{figure}

For this action, participants were asked how they would hide or collapse a set of bullet points underneath a title. \autoref{fig:hide} shows this prompt as it was presented to participants. In pilot testing we found the some people had some difficulty understanding this concept, but in the final study participants seemed very confident in understanding this action, referring to a variety of different situations that had similar actions, like expanding and collapsing text in a wiki or in the note-taking application Notion, opening and closing `drawers', or hiding columns or rows in spreadsheets.

Many participants first responded by saying that for this kind of action that there must be some kind of icon they could tap to hide or show the content, for instance a plus icon or a caret. In this case, we prompted participants to consider how they might define what content they wanted to hide.

The most prevalent gesture for this action was some kind of tapping gesture, though some participants tapped on the title and others tapped on the content they were trying to hide. Another common gesture was to drag the content onto the title, which several participants send was similar to dragging files into a folder on a desktop.
A few participants mentioned that the way the content was created would be key to allowing them to hide it later. We come back to this idea of the implications object creation and representation in the discussion section.

\subsubsection{Delete}

\begin{figure}
  \centering
  \includegraphics[width=.6\linewidth]{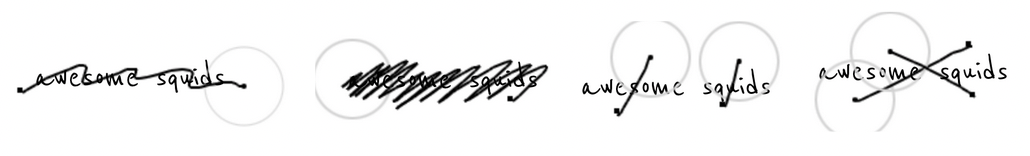}
  \caption{Examples of various ways participants used a `mark' to indicate they wanted to delete the text.}
  \Description{A woman and a girl in white dresses sit in an open car.}
  \label{fig:delete}
\end{figure}

Delete did not have much consensus about the gesture to be used. Many participants make some kind of obscuring mark over the text, whether it be scribbling over the text or drawing an X or something similar. \autoref{fig:delete} shows a variety of marks participants made to indicate delete. Many participants wanted to strike through the text from right to left, and noted that the directionality was important as they expected swiping from left to right would indicate select. These participants often mentioned how striking from right to left evoked the way a cursor would move from right to left when they hit the delete button on a keyboard.
Though there was much variation in the delete gestures, there was clearly a common theme of somehow obscuring the text. 

\subsection{Select Gesture Correlates}
\label{sec:select-correlates}

Several of the actions we studied could reasonably require a select gesture prior to a gesture specific to the action. For instance, to copy text participants often first selected the text with one gesture and then performed another gesture to initiate a copy. This allowed us to look at the consistency of participants' select gestures across different actions, and investigate if participants' select gestures had any correlation with the demographic information we collected.

\begin{figure}
     \centering
     \begin{subfigure}{0.64\textwidth}
         \centering
         \includegraphics[width=\textwidth]{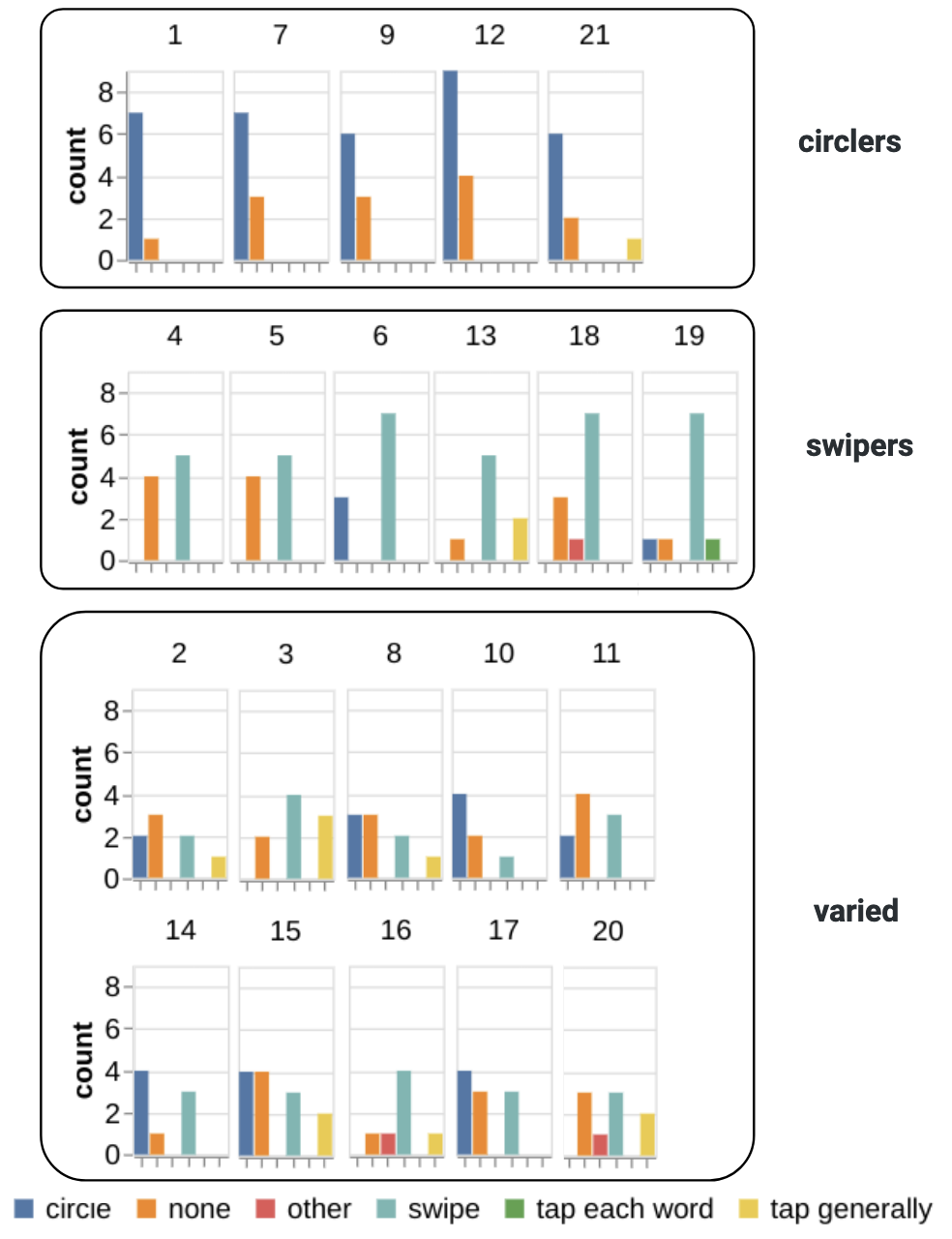}
         \caption{Histograms for each participant's use of select gestures, grouped manually by their observed preferences. While some participants had clear preferences for a `circle' gesture, or a `swipe', most participants used a variety of gestures throughout the study to select ink.}
         \label{fig:select}
     \end{subfigure}
     \begin{subfigure}{0.35\textwidth}
         \centering
         \includegraphics[width=\textwidth]{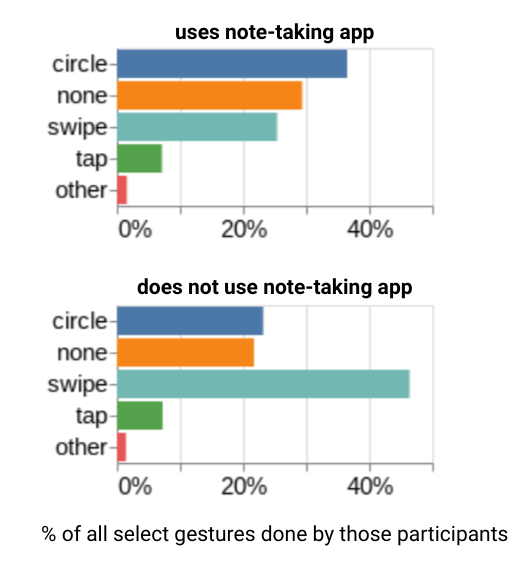}
         \caption{Breakdown of select gestures by whether or not a participant reported using a stylus-native note-taking application. Participants who did not report using such an application were much more likely to swipe to select as compared to their counterparts.}
         \label{fig:select-var}
     \end{subfigure}
        \caption{Because some of the actions we used to prompt participants could involve a select gesture, we were able to look at participants preferences for select gestures across all actions. Participants' preferences for select gestures may be explained be whether they had experience with existing stylus-native note-taking applications.}
        \label{fig:select graphs}
\end{figure}

\autoref{fig:select} shows the prevalence of the different select gestures used broken down by participant. Participants are grouped by the type of select gesture they used. Five participants consistently used the circle gesture to select, or used no select gesture at all (`none'). (A participant might not use a select gesture, for instance, when deleting.) Six participants consistently used a swipe (or `line') gesture, and the remaining were not consistent in the select gesture they used.

We investigated if any of the demographic data we collected had any correlation with the gesture a participant was likely to use. Only previous experience with a note-taking application showed any difference. \autoref{fig:select-var} shows this difference: participants who didn't report using stylus-based note-taking applications were far more likely to use the swipe gesture for select. This doesn't mean that participants who use a note-taking app never used a swipe gesture, but rather that they also relied heavily on circling. We investigate potential reasons for this in the next section.

\subsection{Mental Models}

One value of running this study with a facilitator was that we were able to collect qualitative data about how participants were thinking about their gestures. This allowed us to investigate the mental models and other priors participants were bringing to the elicitation study.

\subsubsection{Word processor.}

Many participants were very attached to the idea of a cursor and keyboard, even when reminded to imagine that all text had been handwritten.
For instance, P10 said, “I would long press and then select. You have to take the cursor from the front to the back.” 
and P4 said, “I would put the cursor here. And then if there was a gesture for a return or something.” These quotes are emblematic of a larger theme where participants expected even handwritten text to have the affordances of a word processor.

This possibly explains participants' preferences for a swipe select gesture versus a circling one as seen in \autoref{fig:select-var}. Participants with less experience with stylus-based note-taking apps may be drawing more heavily on a word processor mental model, where swiping over text is a analogous to dragging a cursor over text. We come back to implications of the word processor mental model in the discussion.

\subsubsection{Annotations as commands}

\begin{figure}
  \centering
  \includegraphics[width=.6\linewidth]{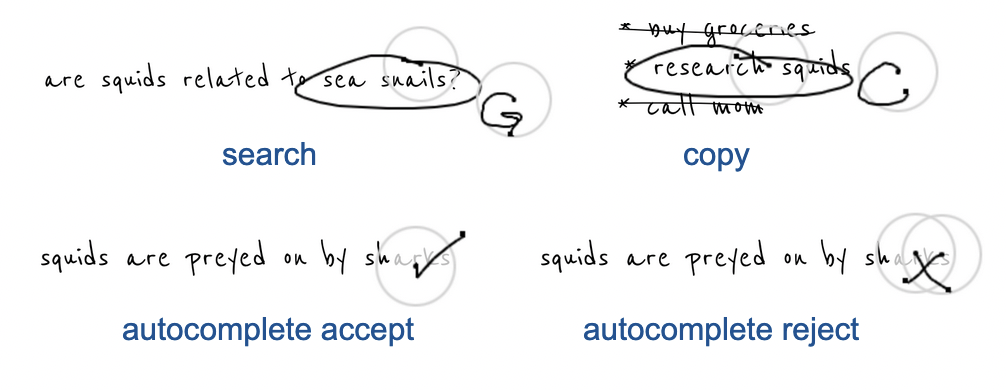}
  \caption{Examples of the various ways participants used annotations to indicate what action they wanted to perform.}
  \Description{A woman and a girl in white dresses sit in an open car.}
  \label{fig:annotations}
\end{figure}

Some gestures involve direct manipulation, and are analogous to interacting with physical objects, like dragging or tapping on text.
But there were many gestures that were more like annotations, where one might imagine an assistant interpreting and acting on the annotations. \autoref{fig:annotations} shows a variety of examples of gestures that participants' performed that were coded as `annotate'.

Interpreting annotations requires distinguishing between types of ink: ink that is drawn as part of a note, and ink meant to be an instruction. We might consider this `permanent' versus `transient' ink. Distinguishing between these types of ink may prove difficult because it relies on understanding the users' intention. Consider a scribbling gesture for delete. How does an intelligent user interface differentiate between a user wanting to remove ink from the screen by scribbling over it versus a user who wants to scribble over ink on the screen but doesn't actually want the ink to disappear?

\subsubsection{Creation priors}

Some participants were sensitive to how ink was created. For instance, P1 used the example of a mindmap: if they drew a connecting line between two words, at that point they would know if the line was ‘attached’ to the words or just stray ink. This would change how they would expect the ink to behave in later actions. Similarly, when considering select gestures, P7 said “I might double tap on it because sometimes that'll make a selection depending on how the word was created.” 

These examples show us how participants expect ink on the screen to have different affordances depending on what occurred when that ink was created. Given that in our study participants were shown ink on the screen instead of being asked to produce this ink themselves, this led to some uncertainty in the participants about how they might be able to perform different actions. We investigate the kind of feedback the participants might have been expecting in order to understand the affordances the ink would have in the discussion section.

\section{Discussion}

\subsection{The Word Processor Legacy}

It is unclear if the reliance on the word processor mental model was somehow an artifact of our study, despite our efforts to have participants imagine a  non-word processor environment, or if it the word processor mental model is simply an extremely prevalent prior. It could be due to the actions we studied, most of which have clear, word processor counterparts. But it seems equally likely that the word processor mental model has pervaded our sense of what computing is. Much computing is simply a more static version of a word processor -- typed text laid out linearly that can be highlighted and copied.

We envision a future of computing that is much more dynamic. Paper has many affordances that the traditional word processor does not, and we believe that future computing interfaces will have some of those affordances. Moving to the more freeform nature of pen and paper will require users to let go of the word processor model, in order to allow for more non-linear affordances. The use of swiping gestures to navigate mobile devices may be a precursor (no pun intended) to this.

\subsection{Feedback at Time of Ink Creation}

Several participants mentioned the kind of feedback they’d expect from the application, to confirm that it would perform their intended action. For instance, P13, when explaining their gesture for `copy', said “Maybe use the stylus to underline or go across the words and then it shows a lift off the screen and I can just move it.” Others commented that “depending on how it was created“ they would expect different gestures to be available. While this is likely a mental model that participants are drawing on from existing applications, it is an important one to consider. When, for instance, is the application correctly interpreting their handwriting? Many participants commented that existing handwriting recognition technology does not work for their handwriting or that even though they have not tried handwriting recognition they imagine that because their handwriting is so poor it would likely not work for them. (On this note, we may note that almost all participants commented on the poor quality of their handwriting, which perhaps suggests that everyone believes their handwriting is below the mean.)

Investigating the kind of feedback users require at the time of ink creation will be an important part of designing intelligent note taking applications. For some actions the feedback is clear, for instance participants will know immediately if their `delete' gesture deletes what they intended, because the ink will or will not disappear. Similarly with `search', participants will either see search results appear or not. However for other actions like `copy', or actions we didn't study but expect to be part of intelligent user interfaces of the future like creating freeform diagrams and mindmaps, it is less clear what kind of feedback will provide users with the necessary information but not distract them from the paper-like experience we expect these applications to want to mimic.

\subsection{Limitations and Future Work}

In this study, participants were asked to imagine how to do something, sometimes against their instincts -- even if they wanted to use a menu option, they were prompted to imagine another way to perform the action. Sometimes they noted a given gesture might not be the best, but it was all they could think of. Some studies, like \cite{wobbrock2009user}, ask participants to rank how well a gesture fits with its action, and how easy it is to perform a gesture. Others, like \cite{nacenta2013memorability}, look at the memorability of new gestures. Future work should investigate the fit and ease of novel gestures, as how likely users are to remember these gestures after being introduced to them.

Additionally, it is important to consider the context in which this elicitation study occurs. The affordances styluses have and the state of note-taking applications are constantly changing. This study represents a moment in time, and a particular, American- and Anglo-centric cultural context. This is not to devalue elicitation studies, nor the elicitation study we present in this paper. Rather, we acknowledge the limits of the study as being extremely contextual. Elicitation studies should be replicated and re-run in order to understand how people's priors vary across cultures and time.

\section{Conclusion}

In this work we studied the gestures users envision performing to enact new actions for an imaginary, intelligent note-taking application. We studied both baseline actions like select and copy as well as novel actions like accepting a handwritten autocomplete and performing an integrated search on handwritten text. Through an elicitation study we found that participants had high agreement for the select and autocomplete accept/reject actions, and much lower agreement on the actions search, copy, hide, and delete. We found that participants with experience using digital note-taking apps are more likely to select by circling text (rather than `highlighting' it) and that many participants still draw on a word processor mental model even when interacting with handwritten digital notes. We discuss how we envision a future of computing that moves away from the word processor mental model and towards the freeform nature of analog pen and paper, where stylus-centric interactions can help people quickly and easily trigger rich writing features.


\bibliographystyle{ACM-Reference-Format}
\bibliography{cite}

\end{document}